# Jaynes-Cummings interaction between low energy free-electrons and cavity photons

Aviv Karnieli[1] and Shanhui Fan[2]

[1]Raymond and Beverly Sackler School of Physics and Astronomy, Tel Aviv University, Ramat Aviv 69978, Tel Aviv, Israel

[2]Department of Electrical Engineering, Stanford University, Stanford, California 94305, USA

**The Jaynes-Cummings Hamiltonian is at the core of cavity quantum electrodynamics, and is ubiquitous in a variety of quantum technologies. The ability to implement and control the various aspects of this Hamiltonian is thus of paramount importance. However, conventional implementations relying on bound-electron systems are fundamentally limited by the Coulomb potential that bounds the electron, in addition to suffering from practical limitations such as requiring cryogenic temperatures for operation and fabrication inhomogeneity. In this work, we propose theoretically a new approach to realize the Jaynes-Cummings Hamiltonian using low energy free-electrons coupled to dielectric microcavities, and exemplify several quantum technologies made possible by this approach. Our approach utilizes quantum recoil, which causes a large detuning that inhibits the emission of multiple consecutive photons, effectively transforming the free-electron into a two-level system coupled to the cavity mode. We show that this approach can be used for generation of single photons with unity efficiency and high fidelity. We then generalize the concept to achieve a multiple-level quantum emitter through a suitable design of cavity modes, allowing for deterministic photon-pair generation and even a quantum SWAP gate between a cavity photon and a free-electron qubit. An increase in coupling strength by a factor of $\sqrt{N}$ can be achieved when an entangled symmetric state of $N$ electrons is used to drive the cavity. Tunable by their kinetic energy and phase-matching to light waves, quantum free-electrons are inherently versatile emitters with an engineerable emission wavelength. As such, they pave the way towards new possibilities for quantum interconnects between photonic platforms at disparate spectral regimes.**

**Introduction**

The Jaynes-Cummings (JC) Hamiltonian[1,2], which describes the resonant interaction of an optical cavity with two electronic energy levels, is quintessential in cavity quantum electrodynamics[3–15] and plays a central role in the development of quantum technology. Many foundational quantum information processing capabilities, including quantum memory[6,7,16], quantum gates[17,18], and deterministic photon sources[19–22], rely upon physical systems described by the JC interaction.

Given its prominence in quantum science and technology, the ability to design and control various aspects of the JC Hamiltonian is of fundamental importance. In the standard implementation in the optical regime, the electronic energy levels are realized in bound electron systems. For atomic and molecular systems[3,4,11], shaping the electronic levels to match a desirable wavelength requires sub-atomic scale control of the atomic potential, posing a serious difficulty. Moreover, for bound electrons in solid-state systems[12,19,20], unavoidable microscopic variations between emitters lead to significant inhomogeneity that hampers many important practical applications.

As opposed to bound-electron systems, free-electrons are fundamentally versatile emitters[23–26] that are dynamically tunable through their kinetic energy, and can thus be tailored to any desired photonic wavelength, from the THz[27] to the x-ray[28] regimes. In the optical domain, free-electron–light interactions[26,29–43] enjoyed drastic advancement in the past decade, demonstrating a plethora of quantum effects[33,41,42,44]. In the context of quantum photonics, it was shown that free-electrons phase-matched to waveguide modes can excite multiple photonic quasiparticles with unity efficiency[34], and even herald the generation of photons in a room-temperature, integrated linear-optical cavity[32,45,46]. However, above keV energies and near the visible range, a free-electron interacting with light is described by a Hamiltonian where the electron forms an infinite energy ladder[33,39–41,47,48]. Such a Hamiltonian is fundamentally different from the JC model.

In this paper, we introduce an alternative and versatile approach to implement the JC Hamiltonian, using sub-keV free-electrons phase-matched to a Fabry-Perot microcavity. We show that the quantum recoil[49–51] experienced by the slow electron following a single photon emission prevents the phase-matching of the electron to a consecutive photon emission. As a result, the JC interaction emerges between the free-electron and the cavity photon, and a certain value of electron-photon coupling then allows for the deterministic, on-demand emission of a single photon with unity efficiency and high-fidelity into the cavity mode.

We then generalize the free-electron JC model to include two possible transitions in a ladder configuration[52], allowing for the fully-deterministic generation of photon pairs. We proceed to show that by considering a lambda configuration, the JC interaction could realize a SWAP operation[17,18] between the state of a photonic qubit encoded in two cavity modes, and the state of a free electron encoded in two energies, which can be used for quantum state transfer between distant cavities and for flying quantum memories[6]. Finally, we show that by using an entangled symmetric state of $N$ free-electrons, the necessary coupling for deterministic photon generation can be reduced by a factor of $\sqrt{N}$, owing to quantum interference of the initial many-electron correlated state[53–56]. Our results pave the way towards deterministic, integrated free-electron quantum-optical sources, and interconnects between photonic quantum information processing units and flying free charged particle qubits[57,58].

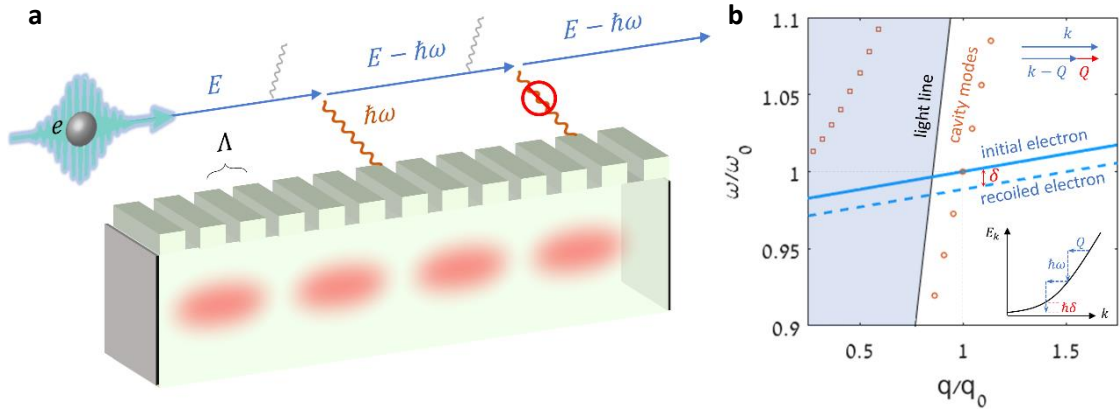

Fig. 1: **concept of deterministic photon generation by slow free electrons**. **a** sub-keV electron of energy $E$ traverses a Fabry-Perot cavity with a periodic corrugation. The periodic grating phase-matches the emission of a photon into the cavity mode and into the free-space continuum (with a far smaller efficiency). Following the first emission, the electron experiences quantum recoil which strongly detunes the emission of a second photon into the cavity. **b** dispersion diagram of the cavity-electron system. The initial electron is phase-matched to a cavity mode $q_0, \omega_0$ below the light line, experiencing a total recoil of $Q = q_0 + 2\pi/\Lambda \cong 2\pi/\Lambda$. Following an emission, the recoiled electron is detuned by $\delta$ from the nearest cavity mode. The solid blue line is defined by energy conservation for the initial electron, $\omega(q) = E(k) - E(k - q - 2\pi/\Lambda)$, passing through a cavity mode, and the dashed line by the energy conservation for the recoiled electron, $\omega(q) = E(k - Q) - E(k - Q - q - 2\pi/\Lambda)$, detuned by $\delta$ from the cavity mode. Upper inset: electron momentum conservation diagram. Lower inset: the effect of the parabolic curvature of the electron dispersion for two consecutive emissions. While the first emission conserves momentum and energy, the second one is detuned.

**Theoretical model**

Our model system comprises a Fabry-Perot resonator of length $L$ with a periodic grating corrugation of periodicity $\Lambda$ of the order of tens of nanometers (Fig. 1a). The periodicity is chosen to be significantly smaller than the free-space wavelength of the resonant modes, ensuring that the cavity modes of interest do not couple to free-space. The grating enables phase-matching of a sub-keV free-electron wavefunction to the cavity modes and to a continuum of free-space modes, in the form of Smith-Purcell (SP) radiation[59]. The free-electron coupling to free-space modes is assumed to be far less efficient than to the cavity mode. This assumption is backed by experimental evidence, demonstrating strong coupling between free-electrons and waveguide polaritons that can reach unity efficiency[34], while the typical efficiency of free-space SP radiation lies between $10^{-4}$ to $10^{-3}$,[60]. Thus, in the forthcoming analytical treatment we shall neglect free-space emission, and later incorporate it as a loss mechanism in our numerical simulations. Under these assumptions, the interaction picture Hamiltonian reads (see Supplementary Material, Section S1 for derivation)

$$H_I = i\hbar \sum_n \int dq \int dk \left[ g_{q,n} e^{i\left(\frac{E_{k+q}-E_k}{\hbar}-\omega_n\right)t} c^\dagger_{k+q} c_k a_n - g^*_{q,n} e^{-i\left(\frac{E_k-E_{k-q}}{\hbar}-\omega_n\right)t} c^\dagger_{k-q} c_k a^\dagger_n \right],$$
(1)

where $E_k$ is the energy of a free-electron with a velocity $v$ and a wavenumber $k$ propagating along $z$, $c_k(c^\dagger_k)$ are the free-electron annihilation (creation) operator at momentum $k$, $a_n$ are bosonic cavity ladder operators of the $n$-th mode, having frequency $\omega_n$ and wavenumber $q_n$, and $g_{q,n} = (ev/\hbar\omega_n) \int d^2\mathbf{r}_T \int dz e^{-iqz} |\psi_T(\mathbf{r}_T)|^2 \mathcal{E}_{n,z}(\mathbf{r}_T, z)$ is the electron-photon coupling coefficient to mode $n$ and with recoil $q$, where $\mathcal{E}_n(\mathbf{r})$ are the vacuum field amplitudes of mode $n$, and $\psi_T(\mathbf{r}_T)$ is the transverse part of the electron wavefunction. In the Hamiltonian Eq. (1), we assumed a transversely-focused free-electron, and further assume that the cavity lifetime is much larger than the interaction time $T = L/v$ of the free-electron with the cavity field.

We now argue that a combination of phase-matching and quantum recoil can transform the Hamiltonian of Eq. (1) to the Jaynes-Cummings form. In general, phase-matching – or momentum conservation – ensures that for an excitation of cavity mode $n$, the electron recoil $q$ equals the mode wavenumber $q_n$, up to a multiple of $2\pi/\Lambda$, owing to the periodic grating coupler. Next, as an example, we choose a specific cavity mode $n_0$, and an initial electron wavevector $k_0$, and denote $Q_{n_0} = q_{n_0} + 2\pi/\Lambda$ as the total electron recoil, such that the emission of a single photon to this mode conserves energy, i.e. $E_{k_0} - E_{k_0-Q_{n_0}} - \hbar\omega_{n_0} = 0$. Following this single-photon emission event, we then consider a subsequent single-photon emission event to this cavity mode. Due to the recoil, this subsequent transition has a non-zero frequency detuning $\delta \equiv \left|(E_{k_0-Q_{n_0}} - E_{k_0-2Q_{n_0}})/\hbar - \omega_{n_0}\right| = \hbar Q_{n_0}^2/m \approx \hbar/m(2\pi/\Lambda)^2$, where we assume a non-relativistic electron, and note that $q_{n_0} \ll 2\pi/\Lambda$. The effect of detuning is measured by the dimensionless quantity $\delta \cdot T$, corresponding to the amount of dynamic phase oscillations during the electron-cavity interaction time $T$. For a swift electron, of keV energies and above, $\delta \cdot T \ll 1$, and the effect of detuning is negligible: the electron keeps emitting photons into mode $n_0$. In this case, Eq. (1) recovers the well-known result for the Hamiltonian of the quantum free-electron–light interaction in the zero-recoil limit [3], predicting Poissonian statistics (Fig. 2a).

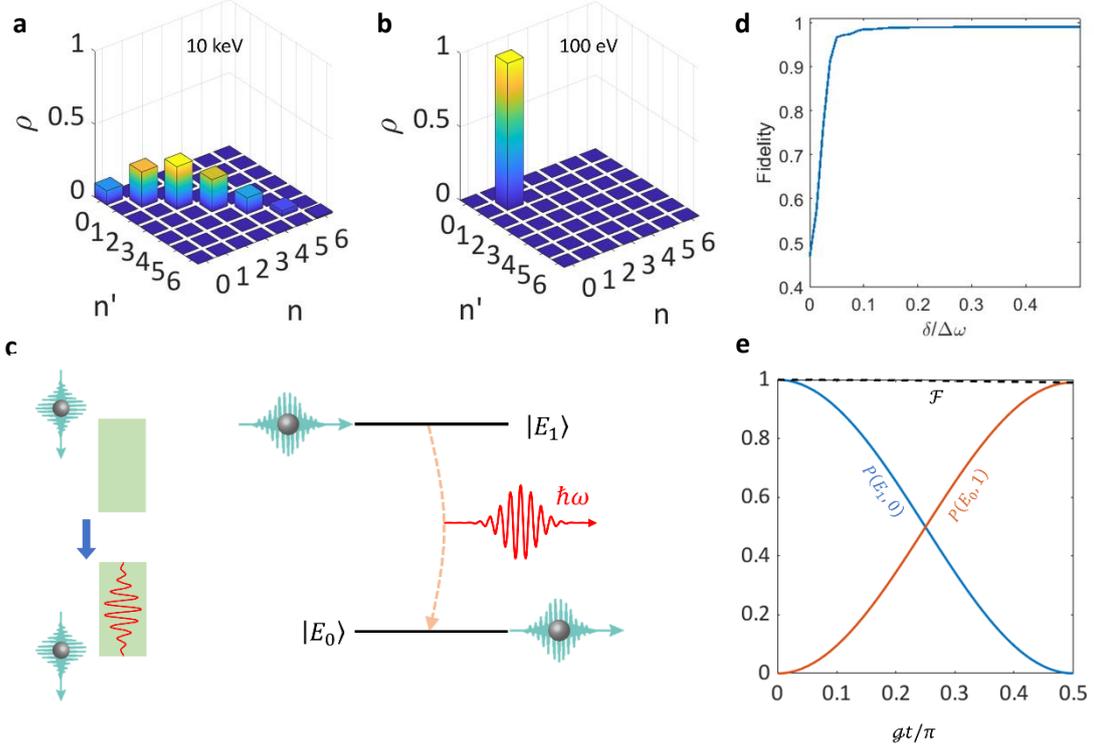

Fig. 2: **Single photon generation by slow free-electrons**. **a** fast electron (above few-keV) interacting with a cavity experiences negligible recoil, and is able to emit multiple photons into the cavity mode, resulting in a Poissonian photon number distribution. **b** pure single-photon state generated by a slow (sub-keV) electron, experiencing significant recoil. **c** Depiction of a single-photon generation by an electron traversing an empty cavity, and the equivalent Jaynes-Cummings model. **d** Fidelity of a pure single-photon state and the simulation result, as a function of the detuning fraction $\delta/\Delta\omega$. **e** probabilities of the joint electron-photon states $|E_1, 0\rangle$ and $|E_0, 1\rangle$, as a function of the quantum coupling $g_Q = gt$, showing the expected Rabi oscillation. Dashed line: fidelity between the numerically calculated state and the analytically predicted state, Eq. (3). The simulation parameters are $g_Q = gL/v = \pi/2$, $L = 10\mu m$, $\lambda_0 = 532nm$, $n = 1.5$, $E = 100eV$ with energy uncertainty of 10meV, free spectral range $\Delta\omega = 2\pi \times 13$ THz $= 55 meV/\hbar$, and free-space emission probability of $10^{-2}$.

However, for a slow electron, the recoil may induce a considerable detuning: either with respect to the original cavity mode $n_0$, or with respect to neighboring modes, $n_0 \pm 1$. This detuning can dramatically alter the emission mechanism, reshaping the photon statistics into that of a single-photon Fock state (Fig. 2b). For this to occur, we require that the smallest detuning from *the nearest* available mode following recoil $\delta_{\min}$, will be large enough to prevent a second photon emission, which corresponds to the requirement $\delta_{\min} T \gg 2\pi$. Since this minimal detuning $\delta_{\min}$ is a fraction $p \leq 1/2$ of the free spectral range $\Delta\omega = \pi c/nL$, the latter condition is independent of cavity length and can be readily fulfilled for slow electrons having $\beta = v/c \ll 1$, since $\delta_{\min} T = (p\pi/n)\beta^{-1}$. In this limit, the rotating wave approximation reduces the Hamiltonian of Eq. (1) to (see Supplementary Material, Section S3 for derivation)

$$H_{2\text{lvl}} = i\hbar g \sigma_+ a - i\hbar g^* \sigma_- a^\dagger, \quad (2)$$

where we dropped the mode number subscript $n_0$, and defined $\sigma_- = c^\dagger_{k_0 - Q_{n_0}} c_{k_0}$ and $\sigma_+ = c^\dagger_{k_0} c_{k_0 - Q_{n_0}}$ as effective two-level system lowering and raising operators, respectively. Namely, the Hamiltonian $H_{2\text{lvl}}$ couples only two possible free-electron momentum states to the desired cavity mode, as illustrated in Fig. 2c.

Interestingly, Eq. (2) is a manifestation of the JC model in the strong-coupling regime. However, as opposed to a bound electron system constituting the matter part, here it consists of a slow free-electron (with a potentially continuous energy spectrum) coupled to an optical field. Unlike the bound electron system, where the transition frequency is set by the atomic potential that bounds the electron, here the transition frequency is set by the resonant frequency of the cavities. Since the cavity frequencies can be defined by nano fabrication, our approach enables significant flexibilities in setting the transition frequency. Moreover, with our approach the transition frequency is automatically matched to the cavity frequency. This removes one of the common difficulties in the engineering of strong coupling with bound electron systems, where the emitter transition frequency has to be carefully tuned to match the cavity resonant frequencies[8,11,12].

Similar to the JC model, the Hamiltonian Eq. (2) has polaritonic eigenstates. Denoting the initial electron energy by $E$, the photon energy by $\hbar\omega$, and photon number by $n$, these eigenstates are of the form $(|E, n-1\rangle \pm |E-\hbar\omega, n\rangle)/\sqrt{2}$ with eigenvalues $\pm\hbar g\sqrt{n}$, with the exception of a ground state $|E-\hbar\omega, 0\rangle$ with eigenvalue 0. Below, we will show how this type of dynamics can result in the realization of several quantum technologies based on the JC Hamiltonian, such as deterministic generation of single photons, photon pairs, and a flying quantum memory based on free-electrons.

As a final remark for this section, we note that our work is related to, but distinct from, previous works that have considered the effect of quantum recoil on the free-electron dynamics[49–51,55,56,61,62]. The breakdown of the free-electron infinite energy ladder structure– and the emergence of a finite set of available free-electron energy levels – is predicted to occur when the quantum recoil experienced by the electron becomes significant. This has been theoretically shown for the quantum regime of the free-electron laser in the x-ray[55,56,61], considering relativistic electrons experiencing backwards Compton scattering. Very recently, it was also predicted that sub-keV electrons driven by a semiclassical laser field in a periodic dielectric structure will demonstrate energy-domain self-trapping[62]. In addition, slow electrons were recently shown to give rise to quantum corrections to Smith-Purcell radiation[49–51] and plasmon emission[63–65], enhanced coupling to plasmonic nanoparticles[66] and complete excitation of nanoscale two-level systems[67]. However, the fully-quantum interaction considered here, between slow free-electrons and photonic microcavities (of particular relevance for integrated quantum photonic technologies[68]) remained until now unexplored.

**Deterministic single-photon generation**

For an initially empty cavity, the joint electron-photon quantum state evolves as a strongly-coupled polaritonic system:

$$|\psi\rangle = \cos|g|t \, |E_1, 0\rangle - e^{-i\arg g} \sin|g|t \, |E_0, 1\rangle, \quad (3)$$

where $E_1 = E$ and $E_0 = E - \hbar\omega$. Deterministic single-photon generation is obtained when the dimensionless coupling constant $g_Q = gT$ reaches $\pi/2$, and the resulting photonic state - a pure single-photon state - is depicted in Fig. 2b. To further confirm Eq. (3) predicting quantum two-level dynamics, we numerically solve the full Hamiltonian of Eq. (1), i.e., with a potentially unbound electron energy spectrum and multiple cavity modes (see Supplementary Material, Section S2 for details). in Fig. 2d we plot the fidelity as a function of detuning $\delta$, between a pure single-photon state and the numerical solution, where we consider 3 cavity modes as well as losses due to the electron coupling to free-space modes with probability $p_{\text{loss}} = 10^{-2}$. Clearly, the fidelity increases with increasing fraction of detuning to free spectral range, $\delta/\Delta\omega$, and plateaus at a value exceeding 0.99. In Fig. 2e, we further plot the probabilities of $|E_1, 0\rangle$ and $|E_0, 1\rangle$ calculated from the numerical solution, as well as the fidelity of the numerically calculated state with the analytical result of Eq. (3), as a function of the dimensionless coupling $g_Q$, for $\delta/\Delta\omega = 1/2$ and $p_{\text{loss}} = 10^{-2}$. The numerical solution demonstrates the expected vacuum Rabi oscillation, showing very high fidelity with the analytical prediction.

**Deterministic photon pair generation**

After establishing the two-level system behavior of the dynamics, we can readily generalize the concept to more than two levels. For a three-level system, the ladder configuration (Fig. 3a) can be realized by considering two adjacent cavity modes, $n_0$ and $n_1$, with frequencies $\omega_0$ and $\omega_1$, respectively, and a frequency spacing $\Delta\omega = \omega_0 - \omega_1$ that exactly equals the recoil experienced by the slow electron when emitting a photon into mode $n_0$, i.e., the second transition should have zero detuning $\delta = 0$. In this manner, it is ensured that following one phase-matched emission into mode $n_0$, the electron transition is phase-matched again with mode $n_1$. Note, however, that these two modes need to be *isolated* from other cavity modes, in the sense that the frequency spacing between them and any other cavity mode is much larger than the recoil. This ensures that the electron transition is far-detuned from other modes following two recoils. While this is not naturally achievable with a Fabry-Perot cavity having a relatively constant spacing between modes, one may, for example, use Bragg reflectors as cavity mirrors, having a narrow bandgap containing the two frequencies of interest. Assuming that this condition is fulfilled, we can write

$$H_{3\text{lvl}} = i\hbar g_0 \sigma_{0,+} a_0 - i\hbar g_0^* \sigma_{0,-} a_0^\dagger + i\hbar g_1 \sigma_{1,+} a_1 - i\hbar g_1^* \sigma_{1,-} a_1^\dagger, \qquad (4)$$

where $\sigma_{i,+}(\sigma_{i,-})$ are the raising (lowering) operators between levels $i$ and $i+1$, with $i = 0,1$ denoting the two possible transitions between the three free-electron energy levels, $E_2 = E$, $E_1 = E - \hbar\omega_1$ and $E_0 = E - \hbar\omega_1 - \hbar\omega_0$; $a_i$ and $g_i$ are the ladder operators of mode $i$ and their coupling constants, respectively (see Fig. 3a). According to the theory of two-mode, three-level Jaynes-Cummings model[52], for an initially empty cavity and electron in the upper energy level $|E_2\rangle$, and $g_0 = g_1 \equiv g$ taken real, the quantum state of the system evolves as

$$|\psi\rangle = \cos^2\left(\frac{gt}{\sqrt{2}}\right)|E_2, 0, 0\rangle - \frac{1}{\sqrt{2}}\sin(\sqrt{2}gt)|E_1, 1, 0\rangle + \sin^2\left(\frac{gt}{\sqrt{2}}\right)|E_0, 1, 1\rangle, \qquad (5)$$

Deterministic photon pair emission is obtained for $g = \pi/\sqrt{2}$. To affirm this analytical prediction with numerical simulation, in Fig. 3b we plot the probabilities of $|E_2, 0,0\rangle$, $|E_1, 1,0\rangle$ and $|E_0, 1,1\rangle$, calculated from the numerical solution of Hamiltonian Eq. (1), as well as the fidelity of the numerically calculated state with the analytical result of Eq. (5), as a function of the dimensionless coupling $gt$ with $p_{\text{loss}} = 10^{-2}$. As can be seen, the probabilities demonstrate the expected vacuum Rabi oscillations while the fidelity remains above 0.98.

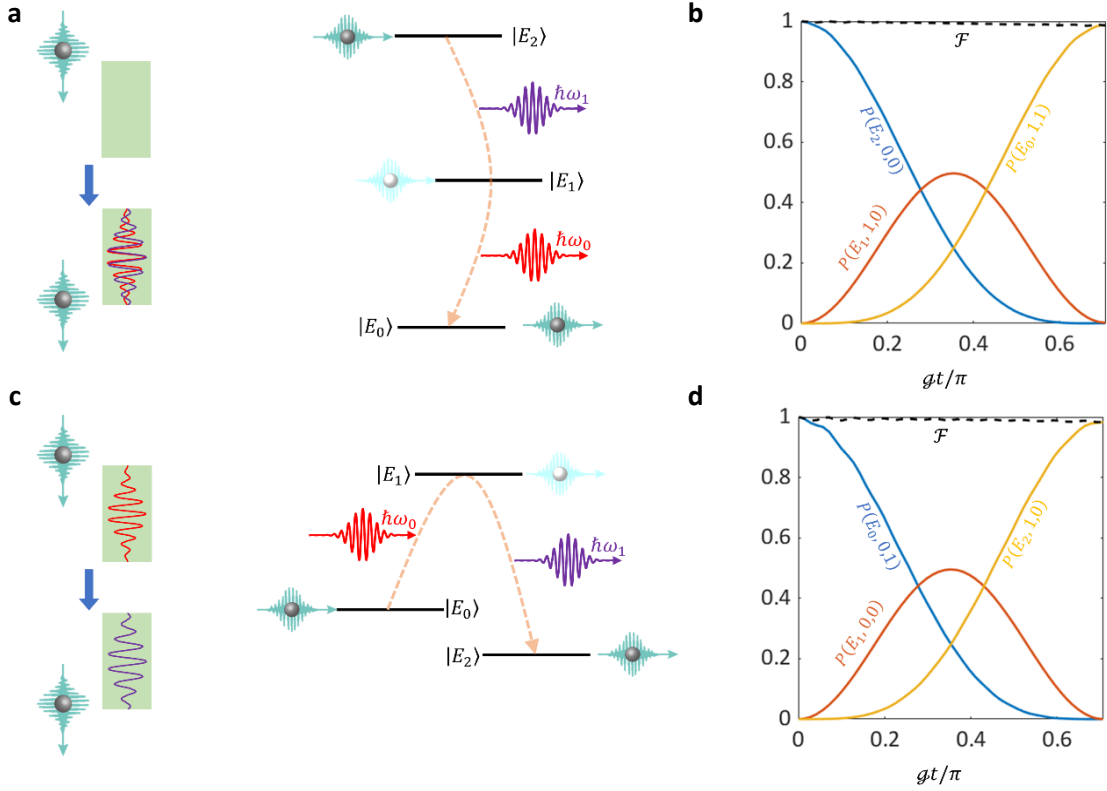

Fig. 3: **Photon pair generation by slow free-electrons, and SWAP operation between a free-electron and a photon. a** Depiction of a slow free-electron emitting a photon pair into two cavity modes, and the corresponding ladder configuration. **b** probabilities of the joint electron-photon states $|E_2, 0,0\rangle, |E_1, 1,0\rangle$ and $|E_0, 1,1\rangle$, as a function of the quantum coupling $g_Q = gt$, showing the expected Rabi oscillations. Dashed line: fidelity between the numerically calculated state and the analytically predicted state, Eq. (5). **c** Depiction of a SWAP operation between a slow free-electron and a photon qubit encoded in two cavity modes, and the corresponding lambda configuration. **d** probabilities of the joint electron-photon states $|E_0, 0,1\rangle, |E_1, 0,0\rangle$ and $|E_2, 1,0\rangle$, as a function of the quantum coupling $g_Q = gt$. Dashed line: fidelity between the numerically calculated state and the analytically predicted state, Eq. (6). The simulation parameters are $g_Q = gL/v = \pi/\sqrt{2}, L = 10\mu m, \lambda_0 = 532nm, n = 1.5, E = 100$eV with energy uncertainty of 10meV, free spectral range $\Delta\omega = 2\pi \times 13$ THz $= 55 meV/\hbar$, and free-space emission probability of $10^{-2}$.

## Photon-electron SWAP gate

Another very useful three-level configuration is that of a $\Lambda$-system, where the phase-matched electron energy levels are a common excited level $E_1 = E$, and two ground states $E_0 = E - \hbar\omega_0$ and $E_2 = E - \hbar\omega_1$, with the respective transitions coupled to modes $n_0$ and $n_1$, as depicted in Fig. (3c). For this to be possible, the electron dispersion line can be chosen to pass through both resonance points $(-q_0, \omega_0)$ and $(q_1, \omega_1)$ at the $q$-$\omega$ plane, considering that each cavity mode $\omega_n$ is a standing wave comprising both positive and negative wavevectors $\pm q_n$. Here, we assume that any other possible electron transitions to adjacent cavity modes are far detuned following a recoil detuning $\delta$. The Hamiltonian takes a similar form to that of Eq. (4). For an initial state $|E_0, 0,1\rangle$, the quantum state of the system evolves as

$$|\psi\rangle = \cos^2\left(\frac{gt}{\sqrt{2}}\right)|E_0, 0,1\rangle + \frac{1}{\sqrt{2}}\sin(\sqrt{2}gt)|E_1, 0,0\rangle - \sin^2\left(\frac{gt}{\sqrt{2}}\right)|E_2, 1,0\rangle, \quad (6)$$

with a similar expression for the evolution of an initial state $|E_2, 1,0\rangle$, where in Eq. (6) one replaces $|E_0, 0,1\rangle$ with $|E_2, 1,0\rangle$ and vice versa. To confirm this analytical prediction with the numerical simulation of Hamiltonian Eq. (1), in Fig. 3d we plot the different state probabilities, and the fidelity between the numerically-calculated quantum state and the analytical result of Eq. (6).

A complete toggle of the free-electron–photon state, from $|E_0, 0,1\rangle$ to $|E_2, 1,0\rangle$, is achieved for $g_Q = gT = \pi/\sqrt{2}$, which realizes the transformations $|E_0, 0,1\rangle \to -|E_2, 1,0\rangle$ and

$|E_2, 1,0\rangle \to -|E_0, 0,1\rangle$. Interestingly, this unitary operation can implement a quantum swap gate[17,18] (up to elementary single-qubit operations) between a photon qubit, encoded as $|\psi\rangle_{\text{ph}} = \alpha_{\text{ph}}|0,1\rangle + \beta_{\text{ph}}|1,0\rangle$, and a free-electron qubit[57,58], encoded as $|\psi\rangle_{\text{el}} = \alpha_{\text{el}}|E_0\rangle + \beta_{\text{el}}|E_2\rangle$. To understand this, consider first a free-electron initiated in the state $|E_0\rangle$: an initial photon state $|0,1\rangle$ toggles the electron state to $-|E_2\rangle$ while the photonic state changes to $|1,0\rangle$. However, an initial photonic state of $|1,0\rangle$ does not change, and does not toggle the electron state. The opposite happens for an electron initiated in $|E_2\rangle$. Thus, for arbitrary photonic and electronic states, we find $(\alpha_{\text{el}}|E_0\rangle + \beta_{\text{el}}|E_2\rangle)(\alpha_{\text{ph}}|0,1\rangle + \beta_{\text{ph}}|1,0\rangle) \to (-\beta_{\text{ph}}|E_0\rangle + \alpha_{\text{ph}}|E_2\rangle)(\beta_{\text{el}}|0,1\rangle - \alpha_{\text{el}}|1,0\rangle)$, realizing the gate $U = \sigma_{Y,\text{el}}\sigma_{Y,\text{ph}}\text{SWAP}$.

**Boosting deterministic photon generation rate with entangled free electrons**
In this final section, we propose that the deterministic photon generation rate can be accelerated using superradiant, symmetric free-electron states as have been recently predicted for Cherenkov radiation[53] and for the quantum free-electron laser[55,56]. For the two-level system scheme, we generalize the Hamiltonian of Eq. (2) by considering $N$ free-electrons:
$$H_{\text{2lvl},N} = i\hbar g S_+ a - i\hbar g^* S_- a^\dagger, \quad (7)$$
where $S_- = \sum_{i=1}^{N} \sigma_{-,i}$ (and similarly for $S_+$) are the collective lowering and raising operators, where now $\sigma_{-,i}$ is the lowering operator of the $i$-th electron. The electrons are assumed to arrive simultaneously at the cavity, and are distinguishable by their transverse position (the total beam width is assumed to be much smaller than the length scale over which the coupling $g$ varies), while electron-electron interactions are neglected.

Borrowing from the theory of superradiance[69], we call the collective ladder states symmetric states, with the $n$-th excited symmetric state denoted by $|\mathbf{n}\rangle_S$ and generated from the collective ground state $|\mathbf{0}\rangle_S = |E_0 E_0 \ldots E_0\rangle_N$ via $|\mathbf{n}\rangle_S = \sqrt{(N-n)!/N!\,n!}\, S_+^n |\mathbf{0}\rangle_S$. The ladder operators then transform the symmetric states according to $S_- |\mathbf{n}\rangle_S = \sqrt{(N-n+1)n}|\mathbf{n}-\mathbf{1}\rangle_S$, and $S_+ |\mathbf{n}\rangle_S = \sqrt{(N-n)(n+1)}|\mathbf{n}+\mathbf{1}\rangle_S$. Although the eigenbasis of $H_{\text{2lvl},N}$ is now more involved than the JC model of a single electron [Eq. (2)], it still retains the block-diagonal internal structure for a fixed total excitation number, a well-known fact of the so-called Tavis-Cummings model[70]. Specifically, for the single-excitation manifold, there are two collective eigenstates of the form $(|\mathbf{1}\rangle_S|0\rangle \pm |\mathbf{0}\rangle_S|1\rangle)/\sqrt{2}$ with eigenvalues $\pm\sqrt{N}|g|$. This means that for an initial single-excitation symmetric state $|\mathbf{1}\rangle_S$, the quantum state of the total system evolves as
$$|\psi\rangle = \cos(\sqrt{N}|g|t)\,|\mathbf{1}\rangle_S|0\rangle - e^{-i\arg g}\sin(\sqrt{N}|g|t)\,|\mathbf{0}\rangle_S|1\rangle, \quad (8)$$
where now, a deterministic single-photon generation is obtained for $g_Q = gT = \pi/2\sqrt{N}$. This reduction of the necessary coupling by a factor of $\sqrt{N}$ is a significant improvement, but it comes at the cost of generating the initial symmetric state $|\mathbf{1}\rangle_S$, which is an entangled state of $N$ free-electrons of the form
$$|\mathbf{1}\rangle_S = \frac{|E_1 E_0 \ldots E_0\rangle + |E_0 E_1 \ldots E_0\rangle + \cdots + |E_0 E_0 \ldots E_1\rangle}{\sqrt{N}}, \quad (9)$$
The state Eq. (9) can cause a free-electron superradiance mechanism that relies on quantum correlations[69], thus differing from conventional many-electron superradiance[35,54] that relies on and temporal and spatial correlations of the many-electron wavefunction. The generation of multi-electron entangled states in electron microscopes is currently being pursued, where observations of few-electron energy correlations have been recently demonstrated.[71,72]

**Experimental considerations**

From an experimental point of view, phase-matched low-energy free-electron light emission has been demonstrated in SEM[24,73] and even on-chip[74]. Thus, we believe that a proof-of-concept of our proposed scheme can be implemented, for example in low-energy electron microscopes[75]. However, to maximize near-field coupling while preventing the electrons from hitting the sample, it will be necessary to focus the slow-electron wavefunction to the nanoscale over a long grazing distance, posing a major engineering challenge. Possible solutions could come from employing non-diffracting electron beams[76], miniaturization of ponderomotive free-electron guiding[77], and future designs of free-electron cavities[78] on the micrometer-scale.

Approaching the complete inversion for $g_Q = \pi/2$ in dielectric structures is another challenging requirement. Values of $g_Q$ reaching unity were recently observed experimentally using swift free-electrons and polaritonic waveguides[34,79]. All-dielectric, periodic 1D-microcavity structures, such as corrugated slot waveguide cavities[80], fiber Bragg gratings supporting BICs[81], corrugated nanowire cavities[82,83], and photonic crystal nanobeam cavities[84] can be promising platforms for reaching this regime in dielectrics, and so does employing special types of photonic resonances[85]. Strong coupling between the free-electron and the cavity can be reached whenever $g \gg \gamma$, where $\gamma$ is the cavity photon loss rate. For $g_Q = \pi/2$, the cavity quality factor $Q = \omega/2\gamma$ needs to satisfy $Q \gg (2L/\lambda)\beta^{-1}$. For the parameters considered in this work, this corresponds to quality factors larger than $10^3$, a fairly reasonable value. The required small energy uncertainty of the electron, on the tens of meV range, can be obtained by state-of-the-art electron monochromators[86].

**Discussion**

We presented a new approach for the realization of the Jaynes-Cummings model using low-energy free-electrons coupled to microcavities. We showed how the quantum recoil experienced by a slow free-electron following a single-photon emission can be harnessed to control its detuning to a consecutive emission. This allows for the engineering of discrete energy levels for the free-electron, and the emergence of a Jaynes-Cummings interaction with the cavity-mode. We then proposed to use this effect for the realization of quantum technologies based on the JC Hamiltonian, such as deterministic single-photon and even photon-pair sources, as well as a flying quantum memory based on a quantum SWAP gate between a free-electron and a photon qubit. The strong coupling was then shown to increase by a factor of $\sqrt{N}$ when an entangled symmetric state of $N$ electrons is used to drive the cavity.

As an outlook for this work, we emphasize that unlike existing bound-electron emitters, free-electrons enable fundamental versatility as they are tunable through their kinetic energy. In this manner, they can emit at different wavelengths when their velocity is changed and when they are phase-matched to different photonic cavities. This trait of free electrons has been already utilized in the few-to-hundred keV regime[23–25], allowing for tailored and versatile light sources. In addition, free electrons have recently shown potential for manipulation and readout of quantum information from matter qubits[87–89], where the use of slow electrons can dramatically increase the electron-matter coupling[67]. Employing the versatility of low-energy electrons for quantum applications can pave the way towards new quantum interconnects between photonic platforms at disparate spectral regimes, and even between matter qubits.

**References**

1. Jaynes, E. T. & Cummings, F. W. Comparison of Quantum and Semiclassical Radiation Theories with Application to the Beam Maser. *Proc. IEEE* **51**, 89–109 (1963).

2. Shore, B.W., Knioght, P.L. The Jaynes-Cummings Model. *J. Mod. Opt.* **40**, 1195 (1993).
3. Kaluzny, Y., Goy, P., Gross, M., Raimond, J. M. & Haroche, S. Observation of Self-Induced Rabi Oscillations in Two-Level Atoms Excited Inside a Resonant Cavity: The Ringing Regime of Superradiance. *Phys. Rev. Lett.* **51**, 1175 (1983).
4. Rempe, G., Walther, H. & Klein, N. Observation of quantum collapse and revival in a one-atom maser. *Phys. Rev. Lett.* **58**, 353 (1987).
5. Liu, W.-Y., Zheng, D.-N. & Hartmann, M. J. Quantum simulation with interacting photons. *J. Opt.* **18**, 104005 (2016).
6. Reiserer, A. & Rempe, G. Cavity-based quantum networks with single atoms and optical photons. *Rev. Mod. Phys.* **87**, 1379–1418 (2015).
7. Specht, H. P. *et al.* A single-atom quantum memory. *Nat. 2011 4737346* **473**, 190–193 (2011).
8. Thompson, R. J., Rempe, G. & Kimble, H. J. Observation of normal-mode splitting for an atom in an optical cavity. *Phys. Rev. Lett.* **68**, 1132 (1992).
9. Meschede, D., Walther, H. & M̈ller, G. One-Atom Maser. *Phys. Rev. Lett.* **54**, 551 (1985).
10. Walther, H., Varcoe, B. T. H., Englert, B. G. & Becker, T. Cavity quantum electrodynamics. *Reports Prog. Phys.* **69**, 1325 (2006).
11. Chikkaraddy, R. *et al.* Single-molecule strong coupling at room temperature in plasmonic nanocavities. *Nat. 2016 5357610* **535**, 127–130 (2016).
12. Reithmaier, J. P. *et al.* Strong coupling in a single quantum dot–semiconductor microcavity system. *Nat. 2004 4327014* **432**, 197–200 (2004).
13. Hacker, B., Welte, S., Rempe, G. & Ritter, S. A photon–photon quantum gate based on a single atom in an optical resonator. *Nat. 2016 5367615* **536**, 193–196 (2016).
14. Frisk Kockum, A., Miranowicz, A., De Liberato, S., Savasta, S. & Nori, F. Ultrastrong coupling between light and matter. *Nat. Rev. Phys. 2019 11* **1**, 19–40 (2019).
15. Duan, L. M. & Kimble, H. J. Scalable photonic quantum computation through cavity-assisted interactions. *Phys. Rev. Lett.* **92**, 127902 (2004).
16. Lvovsky, A. I., Sanders, B. C. & Tittel, W. Optical quantum memory. *Nat. Photonics 2009 312* **3**, 706–714 (2009).
17. Bechler, O. *et al.* A passive photon–atom qubit swap operation. *Nat. Phys.* **14**, 996–1000 (2018).
18. Koshino, K., Ishizaka, S. & Nakamura, Y. Deterministic photon-photon √sWAP gate using a Λ system. *Phys. Rev. A* **82**, (2010).
19. Zhi Xia, S. *et al.* Engineered quantum dot single-photon sources. *Reports Prog. Phys.* **75**, 126503 (2012).
20. Senellart, P., Solomon, G. & White, A. High-performance semiconductor quantum-dot single-photon sources. *Nat. Nanotechnol.* **12**, 1026–1039 (2017).
21. Lounis, B. & Moerner, W. E. Single photons on demand from a single molecule at room temperature. *Nat. 2000 4076803* **407**, 491–493 (2000).
22. Eisaman, M. D., Fan, J., Migdall, A. & Polyakov, S. V. Invited Review Article: Single-photon sources and detectors. *Rev. Sci. Instrum.* **82**, 071101 (2011).
23. Adamo, G. *et al.* Light well: A tunable free-electron light source on a chip. *Phys. Rev. Lett.* **103**, 113901 (2009).
24. Charles Roques-Carmes, Steven E. Kooi, Yi Yang, Aviram Massuda, Phillip D. Keathley, Aun Zaidi, Yujia Yang, John D. Joannopoulos, Karl K. Berggren, I. K. & M. S. Towards integrated tunable all-silicon free-electron light sources. *Nat. Commun.* **10**, 3176 (2019).
25. Karnieli, A. *et al.* Cylindrical Metalens for Generation and Focusing of Free-Electron Radiation. *Nano Lett.* **22**, 5641–5650 (2022).
26. Roques-Carmes, C. *et al.* Free-electron-light interactions in nanophotonics. (2022)

27. Korbly, S. E., Kesar, A. S., Sirigiri, J. R. & Temkin, R. J. Observation of Frequency-Locked Coherent Terahertz Smith-Purcell Radiation. doi:10.1103/PhysRevLett.94.054803.
28. Shentcis, M. *et al.* Tunable free-electron X-ray radiation from van der Waals materials. *Nat. Photonics* **14**, 686–692 (2020).
29. García de Abajo, F. J. Optical excitations in electron microscopy. *Rev. Mod. Phys.* **82**, 209–275 (2010).
30. Dahan, R. *et al.* Imprinting the quantum statistics of photons on free electrons. *Science (80-. ).* **373**, (2021).
31. Polman, A., Kociak, M. & García de Abajo, F. J. Electron-beam spectroscopy for nanophotonics. *Nat. Mater.* **18**, 1158–1171 (2019).
32. Feist, A. *et al.* Cavity-mediated electron-photon pairs. *Science (80-. ).* **377**, 777–780 (2022).
33. Di Giulio, V., Kociak, M. & de Abajo, F. J. G. Probing quantum optical excitations with fast electrons. *Optica* **6**, 1524 (2019).
34. Adiv, Y. *et al.* Observation of 2D Cherenkov radiation and its Quantized Photonic Nature Using Free-Electrons. *Conf. Lasers Electro-Optics (2021), Pap. FM1L.6* FM1L.6 (2021) doi:10.1364/CLEO_QELS.2021.FM1L.6.
35. Gover, A. *et al.* Superradiant and stimulated-superradiant emission of bunched electron beams. *Rev. Mod. Phys.* **91**, (2019).
36. Barwick, B., Flannigan, D. J. & Zewail, A. H. Photon-induced near-field electron microscopy. *Nature* **462**, 902–906 (2009).
37. Feist, A. *et al.* Quantum coherent optical phase modulation in an ultrafast transmission electron microscope. *Nature* **521**, 200–203 (2015).
38. Dahan, R. *et al.* Resonant phase-matching between a light wave and a free-electron wavefunction. *Nat. Phys.* **16**, 1123–1131 (2020).
39. Wang, K. *et al.* Coherent interaction between free electrons and a photonic cavity. *Nature* **582**, 50–54 (2020).
40. Kfir, O. *et al.* Controlling free electrons with optical whispering-gallery modes. *Nature* **582**, 46–49 (2020).
41. Kfir, O. Entanglements of Electrons and Cavity Photons in the Strong-Coupling Regime. *Phys. Rev. Lett.* **123**, 103602 (2019).
42. Ben Hayun, A. *et al.* Shaping quantum photonic states using free electrons. *Sci. Adv.* **7**, 4270–4280 (2021).
43. Henke, J.-W. *et al.* Integrated photonics enables continuous-beam electron phase modulation. (2021).
44. Dahan, R. *et al.* Imprinting the quantum statistics of photons on free electrons. *Science (80-. ).* **373**, (2021).
45. Bendaña, X., Polman, A. & García De Abajo, F. J. Single-photon generation by electron beams. *Nano Lett.* **11**, 5099–5103 (2011).
46. Huang, G., Engelsen, N. J., Kfir, O., Ropers, C. & Kippenberg, T. J. Quantum state heralding using photonic integrated circuits with free electrons. (2022) doi:10.48550/arxiv.2206.08098.
47. Barwick, B., Flannigan, D. J. & Zewail, A. H. Photon-induced near-field electron microscopy. *Nature* **462**, 902–906 (2009).
48. Dahan, R. *et al.* Resonant phase-matching between a light wave and a free-electron wavefunction. *Nat. Phys.* 1–9 (2020).
49. Tsesses, S., Bartal, G. & Kaminer, I. Light generation via quantum interaction of electrons with periodic nanostructures. *Phys. Rev. A* **95**, 013832 (2017).
50. Huang, S. *et al.* Quantum Recoil in Free Electron-Driven Spontaneous Emission from

51. Huang, S. *et al.* Quantum recoil in free-electron interactions with atomic lattices. *Nat. Photonics 2023* 1–7 (2023) doi:10.1038/s41566-022-01132-6.
52. Yoo, H. I. & Eberly, J. H. Dynamical theory of an atom with two or three levels interacting with quantized cavity fields. *Phys. Rep.* **118**, 239–337 (1985).
53. Karnieli, A., Rivera, N., Arie, A. & Kaminer, I. Superradiance and Subradiance due to Quantum Interference of Entangled Free Electrons. *Phys. Rev. Lett.* **127**, 060403 (2021).
54. García de Abajo, F. J. & Di Giulio, V. Optical Excitations with Electron Beams: Challenges and Opportunities. *ACS Photonics* **17**, 36 (2021).
55. Kling, P., Giese, E., Carmesin, C. M., Sauerbrey, R. & Schleich, W. P. High-gain quantum free-electron laser: Emergence and exponential gain. *Phys. Rev. A* **99**, 053823 (2019).
56. Kling, P., Giese, E., Carmesin, C. M., Sauerbrey, R. & Schleich, W. P. High-gain quantum free-electron laser: Long-time dynamics and requirements. *Phys. Rev. Res.* **3**, 033232 (2021).
57. Reinhardt, O., Mechel, C., Lynch, M. & Kaminer, I. Free-Electron Qubits. *Ann. Phys.* **533**, 2000254 (2021).
58. Tsarev, M. V., Ryabov, A. & Baum, P. Free-electron qubits and maximum-contrast attosecond pulses via temporal Talbot revivals. *Phys. Rev. Res.* **3**, 043033 (2021).
59. Smith, S. J. & Purcell, E. M. Visible Light from Localized Surface Charges Moving across a Grating. *Phys. Rev.* **92**, 1069–1069 (1953).
60. Yang, Y. *et al.* Maximal spontaneous photon emission and energy loss from free electrons. *Nat. Phys.* **14**, 894–899 (2018).
61. Peter Kling, Enno Giese, Rainer Endrich, Paul Preiss, Roland Sauerbrey, W. P. S. What defines the quantum regime of the free-electron laser? *New J. Phys.* **17**, 123019 (2015).
62. Eldar, M., Pan, Y. & Krüger, M. Self-trapping of slow electrons in the energy domain. (2022) doi:10.48550/arxiv.2209.14850.
63. Cox, J. D. & García De Abajo, F. J. Nonlinear Interactions between Free Electrons and Nanographenes. *Nano Lett.* **20**, 4792–4800 (2020).
64. Ciattoni, A. Quantum interaction of sub-relativistic aloof electrons with mesoscopic samples. (2022) doi:10.48550/arxiv.2211.07448.
65. Garciía De Abajo, F. J. Multiple excitation of confined graphene plasmons by single free electrons. *ACS Nano* **7**, 11409–11419 (2013).
66. Talebi, N. Strong Interaction of Slow Electrons with Near-Field Light Visited from First Principles. *Phys. Rev. Lett.* **125**, 080401 (2020).
67. García De Abajo, F. J., Dias, E. J. C. & Di Giulio, V. Complete Excitation of Discrete Quantum Systems by Single Free Electrons. *Phys. Rev. Lett.* **129**, 093401 (2022).
68. Wang, J., Sciarrino, F., Laing, A. & Thompson, M. G. Integrated photonic quantum technologies. *Nat. Photonics 2019 145* **14**, 273–284 (2019).
69. Gross, M. & Haroche, S. Superradiance: An essay on the theory of collective spontaneous emission. *Phys. Rep.* **93**, 301–396 (1982).
70. Tavis, M. & Cummings, F. W. Exact Solution for an N-Molecule—Radiation-Field Hamiltonian. *Phys. Rev.* **170**, 379 (1968).
71. Meier, S., Heimerl, J. & Hommelhoff, P. Few-electron correlations after ultrafast photoemission from nanometric needle tips. (2022) doi:10.48550/arxiv.2209.11806.
72. Haindl, R. *et al.* Coulomb-correlated few-electron states in a transmission electron microscope beam. (2022) doi:10.48550/arxiv.2209.12300.
73. Massuda, A. *et al.* Smith-Purcell Radiation from Low-Energy Electrons. *ACS Photonics*

**5**, 3513–3518 (2018).

74. Liu, F. *et al.* Integrated Cherenkov radiation emitter eliminating the electron velocity threshold. *Nat. Photonics 2017 115* **11**, 289–292 (2017).
75. Man, K. L. *et al.* Low energy electron microscopy. *Reports Prog. Phys.* **57**, 895 (1994).
76. Grillo, V. *et al.* Generation of nondiffracting electron bessel beams. *Phys. Rev. X* **4**, 011013 (2014).
77. Zimmermann, R., Seidling, M. & Hommelhoff, P. Charged particle guiding and beam splitting with auto-ponderomotive potentials on a chip. *Nat. Commun. 2021 121* **12**, 1–6 (2021).
78. Turchetti, M. *et al.* Design and simulation of a linear electron cavity for quantum electron microscopy. *Ultramicroscopy* **199**, 50–61 (2019).
79. Adiv, Y. *et al.* Observation of 2D Cherenkov Radiation. *Phys. Rev. X* **13**, 011002 (2023).
80. Wang, X. *et al.* Silicon photonic slot waveguide Bragg gratings and resonators. *Opt. Express, Vol. 21, Issue 16, pp. 19029-19039* **21**, 19029–19039 (2013).
81. Gao, X., Zhen, B., Soljačić, M., Chen, H. & Hsu, C. W. Bound States in the Continuum in Fiber Bragg Gratings. *ACS Photonics* **6**, 2996–3002 (2019).
82. Fu, A., Gao, H., Petrov, P. & Yang, P. Widely Tunable Distributed Bragg Reflectors Integrated into Nanowire Waveguides. *Nano Lett.* **15**, 6909–6913 (2015).
83. Pan, Z. W. *et al.* Single-mode guiding properties of subwavelength-diameter silica and silicon wire waveguides. *Opt. Express, Vol. 12, Issue 6, pp. 1025-1035* **12**, 1025–1035 (2004).
84. Hu, S. *et al.* Experimental realization of deep-subwavelength confinement in dielectric optical resonators. *Sci. Adv.* **4**, (2018).
85. Yang, Y. *et al.* Photonic flatband resonances for free-electron radiation. *Nat. 2023 6137942* **613**, 42–47 (2023).
86. Krivanek, O. L. *et al.* Progress in ultrahigh energy resolution EELS. *Ultramicroscopy* **203**, 60–67 (2019).
87. Gover, A. & Yariv, A. Free-Electron-Bound-Electron Resonant Interaction. *Phys. Rev. Lett.* **124**, 064801 (2020).
88. Ruimy, R., Gorlach, A., Mechel, C., Rivera, N. & Kaminer, I. Toward Atomic-Resolution Quantum Measurements with Coherently Shaped Free Electrons. *Phys. Rev. Lett.* **126**, 233403 (2021).
89. Zhao, Z., Sun, X. Q. & Fan, S. Quantum Entanglement and Modulation Enhancement of Free-Electron-Bound-Electron Interaction. *Phys. Rev. Lett.* **126**, 233402 (2021).

# Supplementary Material: Jaynes-Cummings interaction between low energy free-electrons and cavity photons

**Table of contents**

**SI1. Interaction Hamiltonian**

In this section, we derive the interaction-picture Hamiltonian of Eq. (1) of the main text.

We begin with the interaction Hamiltonian in second quantization, given by

$$H_{\text{int}} = \int d^3\mathbf{r}\, \mathbf{j}(\mathbf{r}) \cdot \mathbf{A}(\mathbf{r}), \quad (\text{SI}1.1)$$

where

$$\mathbf{j}(\mathbf{r}) = \frac{e}{2m}\hat{\psi}^\dagger(\mathbf{r})(-i\hbar\boldsymbol{\nabla})\hat{\psi}(\mathbf{r}) - \frac{e}{2m}\left(-i\hbar\boldsymbol{\nabla}\hat{\psi}^\dagger(\mathbf{r})\right)\hat{\psi}(\mathbf{r}), \quad (\text{SI}1.2)$$

is the current density operator for nonrelativistic electrons, with

$$\hat{\psi}(\mathbf{r}) = \int d^3\mathbf{k}\, e^{i\mathbf{k}\cdot\mathbf{r}} c_\mathbf{k}, \quad (\text{SI}1.3)$$

denoting the position-space fermionic annihilation operator, $\{\hat{\psi}(\mathbf{r}), \hat{\psi}^\dagger(\mathbf{r})\} = \delta(\mathbf{r}-\mathbf{r}')$, and $c_\mathbf{k}$ the momentum-space annihilation operator. Finally, the vector potential operator is given by

$$\mathbf{A}(\mathbf{r}) = i\sum_j \frac{1}{\omega_j}\boldsymbol{\mathcal{E}}_j(\mathbf{r})a_j - \frac{1}{\omega_j}\boldsymbol{\mathcal{E}}_j^*(\mathbf{r})a_j^\dagger, \quad (\text{SI}1.4)$$

where $\boldsymbol{\mathcal{E}}_j(\mathbf{r}), a_j$ and $\omega_j$ are the electric field mode envelope, annihilation operator, and frequency, of mode $j$.

Substituting $\psi(\mathbf{r})$ we can express the current density operator in terms of the momentum-space operators as

$$\mathbf{j}(\mathbf{r}) = e\int d^3\mathbf{k}' \int d^3\mathbf{k}\, e^{i(\mathbf{k}-\mathbf{k}')\cdot\mathbf{r}} \frac{\hbar(\mathbf{k}+\mathbf{k}')}{2m} c_{\mathbf{k}'}^\dagger c_\mathbf{k} \cong e\mathbf{v}_0 \int d^3\mathbf{k}' \int d^3\mathbf{k}\, e^{i(\mathbf{k}-\mathbf{k}')\cdot\mathbf{r}} c_{\mathbf{k}'}^\dagger c_\mathbf{k}$$
$$= e\mathbf{v}_0 \hat{\psi}^\dagger(\mathbf{r})\hat{\psi}(\mathbf{r}), \quad (\text{SI}1.5)$$

where we assumed non-relativistic electrons propagating along the $\hat{\mathbf{z}}$ direction (referred to also as the longitudinal direction below), and we approximated $\hbar(\mathbf{k}+\mathbf{k}')/2m \cong \mathbf{v}_0$ as the initial electron group velocity vector parallel to the $\hat{\mathbf{z}}$ direction, and lastly, substituted the inverted relation for the momentum-space annihilation operator in terms of the position space operator as $c_\mathbf{k} = \int d^3\mathbf{r}\, e^{-i\mathbf{k}\cdot\mathbf{r}} \hat{\psi}(\mathbf{r})$. Substituting the expression for the current operator, we find

$$H_{\text{int}} = i\sum_j \frac{ev_0}{\omega_j} \int d^3\mathbf{r}\, \mathcal{E}_{j,z}(\mathbf{r}) \hat{\psi}^\dagger(\mathbf{r})\hat{\psi}(\mathbf{r}) a_j + h.c., \quad (\text{SI}1.6)$$

We would like to reduce Hamiltonian (SI1.6) to a one-dimensional Hamiltonian acting on the longitudinal part of the free-electron state. For this we expand the electron annihilation operator $\hat{\psi}(\mathbf{r})$ in terms of a set of orthonormal modes $\phi_n(\mathbf{r}_T)e^{ikz}$, as

$$\hat{\psi}(\mathbf{r}) = \int dk \sum_n \phi_n(\mathbf{r}_T)e^{ikz} c_{k,n}, \qquad (SI1.7)$$

where $\phi_n(\mathbf{r}_T)$ are transverse wavefunctions, and substitute into SI1.6,

$$H_{\text{int}} = i \sum_j \sum_m \sum_n \frac{ev_0}{\omega_j} \int dk \int dq \int dz \int d^2\mathbf{r}_T\, e^{-iqz} \mathcal{E}_{j,z}(\mathbf{r}_T,z)\phi_m^*(\mathbf{r}_T)\phi_n(\mathbf{r}_T) c_{k+q,m}^\dagger c_{k,n} a_j + h.c., \qquad (SI1.8)$$

As an approximation, we now assume that in the resulting summation, we keep only the terms with $m = n = 0$, i.e. we assume that the electron is in a particular transverse mode $\phi_0(\mathbf{r}_T)$ that does not change upon electron-photon interaction. Suppressing the wavefunction subscripts $m, n$, the Hamiltonian becomes

$$H_{\text{int}} = i \sum_j \frac{ev_0}{\omega_j} \int dk \int dq \int dz \int d^2\mathbf{r}_T\, e^{-iqz} |\phi(\mathbf{r}_T)|^2 \mathcal{E}_{j,z}(\mathbf{r}_T,z) c_{k+q}^\dagger c_k a_j + h.c., \qquad (SI1.9)$$

we define the electron-photon coupling as

$$g_{q,j} = \frac{ev_0}{\hbar\omega_j} \int dz \int d^2\mathbf{r}_T\, e^{-iqz} |\phi(\mathbf{r}_T)|^2 \mathcal{E}_{j,z}(\mathbf{r}_T,z), \qquad (SI1.10)$$

resulting in

$$H_{\text{int}} = i\hbar \sum_j \int dk \int dq\, g_{q,j} c_{k+q}^\dagger c_k a_j + h.c., \qquad (SI1.11)$$

In the interaction picture, this Hamiltonian is given by

$$H_I = i\hbar \sum_j \int dq \int dk \left[ g_{q,j} e^{i\left(\frac{E_{k+q}-E_k}{\hbar}-\omega_j\right)t} c_{k+q}^\dagger c_k a_j - g_{q,j}^* e^{-i\left(\frac{E_k-E_{k-q}}{\hbar}-\omega_j\right)t} c_{k-q}^\dagger c_k a_j^\dagger \right], \qquad (SI1.12)$$

recovering Eq. (1) of the main text. Finally, we can express the generic form of the coupling constant for use in the numerical solution, by approximating the integral over the transverse electron density in Eq. (SI1.10). This results in an averaging of the field over the transverse dimension, e.g.

$$\int d^2\mathbf{r}_T\, |\phi(\mathbf{r}_T)|^2 \mathcal{E}_{j,z}(\mathbf{r}) \cong \langle \mathcal{E}_{j,z} \rangle_T \mathcal{E}_{j,z}(z), \qquad (SI1.13)$$

where

$$\mathcal{E}_{j,z}(z) = \begin{cases} e^{i\left(q_j + \frac{2\pi}{\Lambda}\right)z}, & |z| \leq L/2 \\ 0, & \text{otherwise} \end{cases}, \qquad (SI1.14)$$

such that

$$\mathscr{g}_{j,q} = \frac{ev_0}{\hbar\omega_j}\langle\mathcal{E}_{j,z}\rangle_T \int_{-L/2}^{L/2} e^{i\left(q_j+\frac{2\pi}{\Lambda}-q\right)z} dz = \frac{ev_0}{\hbar\omega_j}\langle\mathcal{E}_{j,z}\rangle_T L \operatorname{sinc}\left[\left(q_j+\frac{2\pi}{\Lambda}-q\right)\frac{L}{2}\right],$$
(SI1.15)

## SI2. Numerical solution

We solve the Schrodinger equation in the interaction picture

$$i\hbar\frac{d}{dt}|\psi\rangle_I = H_I|\psi\rangle_I, \quad \text{(SI2.1)}$$

by considering the general solution

$$|\psi\rangle_I = \sum_{k,\{n_j\}} \psi_{k,n_1,n_2,\dots,n_N}|k,n_1,n_2,\dots,n_N\rangle, \quad \text{(SI2.2)}$$

where $j = 1,2,\dots,N$ are the different electromagnetic modes.

Substituting into the Schrodinger equation, we obtain a set of differential equations for the coefficients $\psi_{k,n_1,n_2,\dots,n_N}$:

$$\dot{\psi}_{k,n_1,n_2,\dots,n_N} = \sum_j \int dq \left[\mathscr{g}_{q,j} e^{i\left(\frac{E_k-E_{k-q}}{\hbar}-\omega_j\right)t}\sqrt{n_j+1}\,\psi_{k-q,n_1,\dots,n_j+1,\dots,n_N}\right.$$
$$\left. - \mathscr{g}_{q,j}^* e^{-i\left(\frac{E_{k+q}-E_k}{\hbar}-\omega_j\right)t}\sqrt{n_j}\,\psi_{k+q,n_1,\dots,n_j-1,\dots,n_N}\right], \quad \text{(SI2.3)}$$

where, for a nonrelativistic electron, we have

$$\frac{E_k - E_{k-q}}{\hbar} = \frac{\hbar}{m}kq - \frac{\hbar q^2}{2m}, \quad \text{(SI2.4a)}$$

$$\frac{E_{k+q} - E_k}{\hbar} = \frac{\hbar}{m}kq + \frac{\hbar q^2}{2m}, \quad \text{(SI2.4b)}$$

The results presented in the paper are done by carrying out the numerical integration of Eq. (SI2.3) taking into account all the coupling coefficients of $\mathscr{g}_{q,j}$. We consider five photonic modes, of which one or two are considered as loss channels to free-space radiation with coupling $\sqrt{p_{\text{loss}}}$. The photon number cutoff in each mode is $N = 3$, and the electron momentum continuous Hilbert space spans $7Q$ where $Q$ is the longitudinal recoil for exciting the cavity.

## SI3. Analytic approximation and requirement on detuning

For $L \gg \Lambda$, $L \operatorname{sinc}\left[\left(q_j+\frac{2\pi}{\Lambda}-q\right)\frac{L}{2}\right] \to \delta\left(q_j+\frac{2\pi}{\Lambda}-q\right)$ and thus, the coupling constant of Eq. (SI1.15) reduces to

$$\mathscr{g}_{j,q} = \frac{ev_0}{\hbar\omega_j}\langle\mathcal{E}_{j,z}\rangle_T \delta(q-Q_j), \quad \text{(SI3.1)}$$

where $Q_j = q_j + \frac{2\pi}{\Lambda}$ is the total recoil. When substituted into the interaction-picture Hamiltonian of Eq. (SI1.12), this results in

$$H_\mathrm{I} = i\hbar \sum_j \int dk \left[ \mathscr{g}_j e^{i\left(\frac{\hbar}{m}kQ_j + \frac{\hbar Q_j^2}{2m} - \omega_j\right)t} c^\dagger_{k+Q_j} c_k a_j - \mathscr{g}_j^* e^{-i\left(\frac{\hbar}{m}kQ_j - \frac{\hbar Q_j^2}{2m} - \omega_j\right)t} c^\dagger_{k-Q_j} c_k a_j^\dagger \right],$$

(SI3.2)

where $\mathscr{g}_j = \frac{ev_0}{\hbar \omega_j} \langle \mathcal{E}_{j,z} \rangle_T$. We now consider phase matching of an initial electron at $k_0$ spontaneously emitting a photon into mode $j_0$ (and ending up with momentum $k_0 - Q_{j_0}$) such that

$$\frac{\hbar}{m} k_0 Q_{j_0} - \frac{\hbar Q_{j_0}^2}{2m} - \omega_{j_0} = 0, \qquad \text{(SI3.3)}$$

Note that this automatically means that the reverse process, i.e., absorption of a single photon from mode $j_0$ by an electron of initial momentum $k_0 - Q_{j_0}$ (ending up with momentum $k_0$) is also phase matched:

$$\frac{\hbar}{m}(k_0 - Q_{j_0})Q_{j_0} + \frac{\hbar Q_{j_0}^2}{2m} - \omega_j = \frac{\hbar}{m} k_0 Q_{j_0} - \frac{\hbar Q_{j_0}^2}{2m} - \omega_{j_0} = 0, \qquad \text{(SI3.4)}$$

We now require that following a single photon emission into mode $j_0$ by an electron with initial momentum $k_0$ and final momentum $k_0 - Q_{j_0}$, the subsequent emission into this mode, or to any other adjacent mode, will be detuned. Similarly, we require that following a single photon absorption from mode $j_0$ by an electron with initial momentum $k_0 - Q_{j_0}$ and final momentum $k_0$ will be also detuned. We denote the detuning from mode $j$ for each of these two cases by

$$\delta_j = \begin{cases} \left| \frac{\hbar}{m}(k_0 - Q_{j_0})Q_j - \frac{\hbar Q_j^2}{2m} - \omega_j \right|, & \text{emission} \\ \left| \frac{\hbar}{m} k_0 Q_j + \frac{\hbar Q_j^2}{2m} - \omega_j \right|, & \text{absorption} \end{cases}, \qquad \text{(SI3.5)}$$

we can now approximate $\omega_j = \omega_{j_0} + (j - j_0)\Delta\omega$ with $\Delta\omega > 0$ denoting the FSR of the cavity, and $Q_j \cong Q_{j_0} \cong 2\pi/\Lambda$. Using Eq. (SI3.3) we write $\delta_j$ as

$$\delta_j = \begin{cases} \left| \frac{\hbar}{m}\left(\frac{2\pi}{\Lambda}\right)^2 + (j - j_0)\Delta\omega \right|, & \text{emission} \\ \left| \frac{\hbar}{m}\left(\frac{2\pi}{\Lambda}\right)^2 - (j - j_0)\Delta\omega \right|, & \text{absorption} \end{cases}, \qquad \text{(SI3.6)}$$

We can now define the minimal detuning to any mode $j$ (including $j_0$) as

$$\delta_\mathrm{min} = \min_j \delta_j = \min_j \left| \frac{\hbar}{2m}\left(\frac{2\pi}{\Lambda}\right)^2 - |j - j_0|\Delta\omega \right|, \qquad \text{(SI3.7)}$$

and require that, during the interaction time $T = L/v$, the accumulated phase mismatch due to detuning will be much larger than $2\pi$, i.e.

$$\delta_\mathrm{min} T \gg 2\pi, \qquad \text{(SI3.8)}$$

We observe that by definition, $\delta_\mathrm{min}$ will always be a fraction $0 \leq p \leq 1/2$ of the FSR, that is

$$\delta_{\min} = p\Delta\omega = p\frac{\pi c}{nL}, \qquad \text{(SI3.9)}$$

Thus, the requirement on the phase mismatch becomes

$$\delta_{\min} T = p\frac{\pi c}{nL} \times \frac{L}{v} = p\frac{\pi}{n\beta}, \qquad \text{(SI3.10)}$$

which readily exceeds $2\pi$ for slow electrons, having $\beta \ll 1$. If this requirement is fulfilled, all the temporal exponentials in Eq. (SI3.2) other than those associated with the transitions $k_0, 0 \to k_0 - Q_{j_0}, 1_{j_0}$ and $k_0 - Q_{j_0}, 1_{j_0} \to k_0, 0$ are fast-oscillating and can be neglected, resulting in the Hamiltonian

$$H_I = i\hbar g_{j_0} c^\dagger_{k_0} c_{k_0-Q_{j_0}} a_{j_0} - i\hbar g^*_{j_0} c^\dagger_{k_0-Q_{j_0}} c_{k_0} a^\dagger_{j_0}$$

which is the JC Hamiltonian. Denoting $g \equiv g_{j_0}$, $\sigma_- \equiv c^\dagger_{k_0-Q_{j_0}} c_{k_0}$, $\sigma_+ \equiv c^\dagger_{k_0} c_{k_0-Q_{j_0}}$ and $a \equiv a_{j_0}$, we recover Eq. (2) of the main text,

$$H_{2\text{lvl}} = i\hbar g \sigma_+ a - i\hbar g^* \sigma_- a^\dagger$$